\documentclass[aps, 10pt, twocolumn,preprintnumbers,amsmath,amssymb]{revtex4-1}

\usepackage{graphicx}
\usepackage{dcolumn}
\usepackage{bm}
\usepackage{color}
\usepackage{braket}

\begin{document}

\title{  Michelson Interferometry  of High-order Harmonic Generation in Solids}

\author{Jian-Zhao Jin$^{1}$}
\thanks{These authors equally contribute to this work.}
\author{Hao Liang$^{1}$}
\thanks{These authors equally contribute to this work.}
\author{Xiang-Ru Xiao$^{1}$}
\author{Mu-Xue Wang$^{1}$}
\author{Si-Ge Chen$^{1}$}
\author{Xiao-Yuan Wu$^{1}$}

\author{Qihuang Gong$^{1,2}$}

\author{Liang-You Peng$^{1,2}$}
\email{\vspace*{0.1cm} Corresponding author: liangyou.peng@pku.edu.cn \vspace*{0.3cm}}
\affiliation{$^{1}$State Key Laboratory for Mesoscopic Physics  and  Collaborative Innovation Center of Quantum Matter, School of Physics,  Peking University, Beijing 100871, China}
\affiliation{$^{2}$Collaborative Innovation Center of Extreme Optics, Shanxi University, Taiyuan, Shanxi 030006, China}

\date{\today}

\begin{abstract}

For the high-order harmonic generation in solids, we find   a distinct and clean interference pattern in the high-energy end of the spectrum which can be interpreted as a Michelson interferometer of the Bloch electron. Our results are  achieved by a numerical solution to the time-dependent Schr\"odinger equation of the quasi-electron in solids and can be   explained by an analytical model based on the principle of the   Michelson interferometry.
   The present study  deepens our understanding of the HHG mechanism in crystalline materials and  may find potential applications in imaging of  the dispersion relation or topological structure of the energy bands in solids.

\end{abstract}

\maketitle

Waveform-controllable short laser pulses have become available in a large range of wavelengths from the x-ray to the THz regime~\cite{RevModPhys.72.545, Hentschel2001, Paul,
PhysRevLett.116.205003,Kampfrath2013}. These new light sources have provided us with the feasibilities to trace or even control many ultrafast dynamics happening in the gas, solid, and liquid phase with a simultaneous high temporal and spatial resolution~\cite{Corkum2007,Krausz, PENG20151,Schubert2014, Schultze2014}.
Very recently,  currents induced by a waveform-controllable few-cycle laser pulse has been observed in  the monolayer graphene~\cite{Higuchi2017a}, which has been interpreted as  repeated Landau-Zener transitions.

As a typical ultrafast nonlinear phenomena, the high-order harmonic generation~(HHG) has attracted  intensive investigations and found many applications in the past three decades~\cite{McPherson1987,1-3-FERRAY1988}.
In very recent years, great experimental and theoretical attention  has been paid to the HHG processes in  solids~\cite{Ghimire2011,Vampa2015b, Luu2015,Ndabashimiye2016, You2017, Sivis00041, Yoshikawa2017,Tamaya2016,JiangLu2018}.  Various light-field-driven effects  have been explored   in both semiconductors~\cite{Schultze2014} and narrow band-gap systems~\cite{Higuchi2017a}. Some studies have shown the possibilities of potential applications of nonlinear ultrafast phenomena to the material sciences and devices~\cite{schiffrin_optical-field-induced_2013,Schultze2013,Vampa2015a,Garg2016,luu_measurement_2018}.

The Michelson interferometer has served as a milestone in physics~\cite{Michelson1887} and has shown its great power in many fields of basic research and practical applications. It can be realized by a photon, an electron or other microscopic particles under many circumstances.
The Michelson interferometer has enabled multiple breakthroughs in  many fields~\cite{Santori2002,Dolling2006, Usenko2017}.
The Landau-Zener transition happens  between two energy levels when the system is swept accross an avoided crossing~\cite{Zener1934}.
It plays vital roles in various quantum systems and has found many important applications~\cite{Sillanpaa2006,Kling2010,zenesini_time-resolved_2009,trompeter_visual_2006}.
During the Landau-Zener tunneling, the phase accumulated between the transitions may result in a constructive or destructive interference~\cite{Shevchenko2010}.  These kinds of interferometry have been applied to various systems~\cite{Ji2003,Oliver2005,trompeter_bloch_2006, Cao2013, Wei2017}.
 Some interferometric methods have been used to measure Berry phases and topological properties of materials~\cite{Abanin2013,Grusdt2016}.

In this Letter, by varying the peak intensity of the  laser pulse, we identify an unusual overall oscillation in the high-energy end of the harmonic spectra {in solids}. Our results are based on the numerical solution to the time-dependent Schr\"odinger equation~(TDSE) of the quasi-electron.
We find  that this oscillation can be regarded as a result of  the Michelson interferometry of the Bloch electron  when it is driven by the vector potential to the region where the Landau-Zener transition can occur between two bands.
The interference fringes  can be nicely reproduced   {by} using an analytical model based on the principle of the Michelson interferometer. Atomic units are adopted unless otherwise stated.

\begin{figure*}[tbp]
\centering
\includegraphics[width=\textwidth]{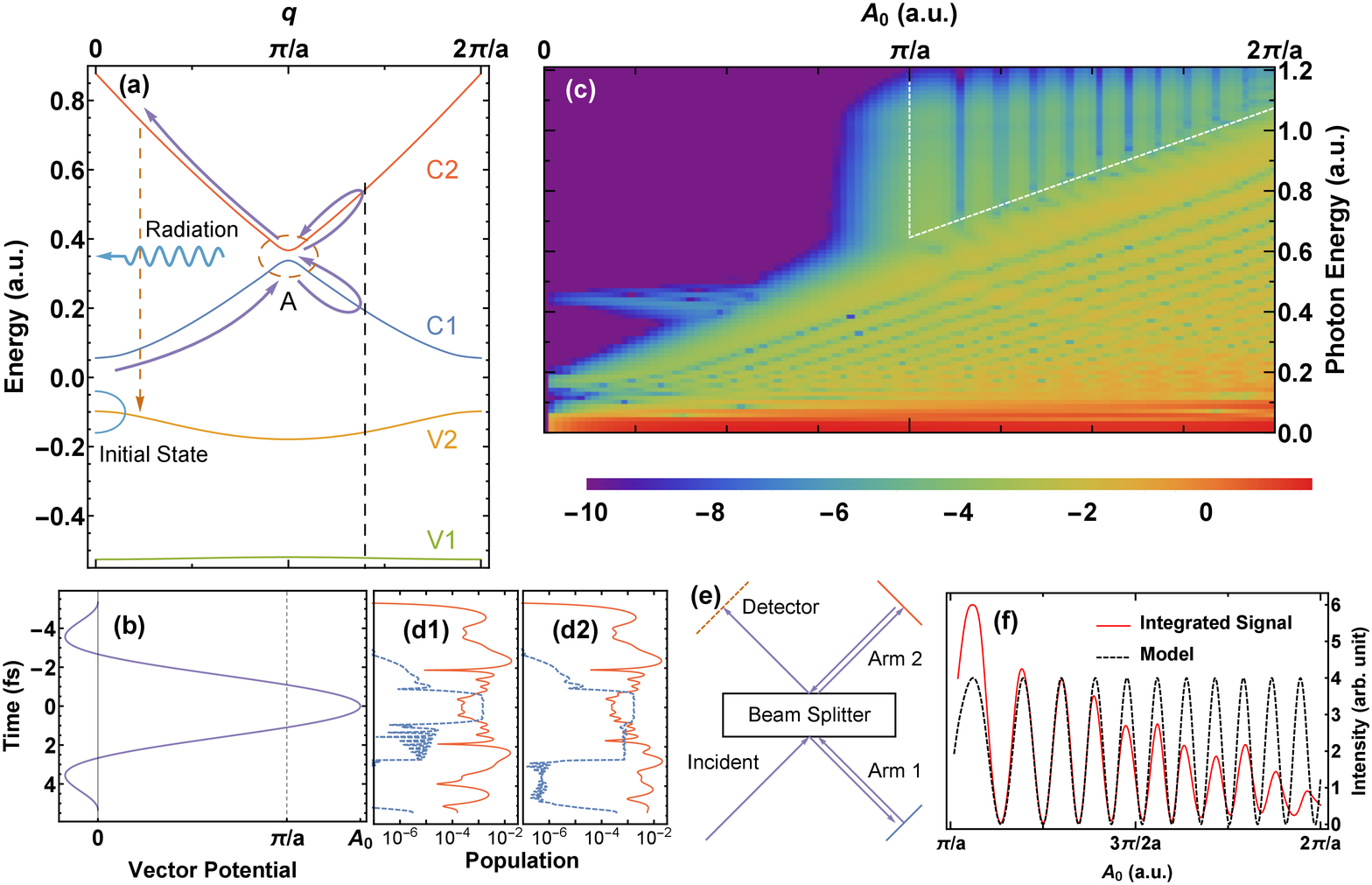}
\caption{The realization of the Michelson interferometer  of Bloch electron in the HHG spectra.   For the model with the first four bands shown in (a), the harmonic spectra  are shown in (c) plotted in the logarithmic scale as a function of  $A_0$ of the pulse, as shown in (b). The Michelson interferometry is present at the top-right corner of (c), enclosed by the white dotted lines. The Bloch electron Michelson interferometer is in the same principle with the usual optical one, as sketched in (e). The interferences minimum and maximum can be verified by checking the populations  of C1 (solid red line) and C2 (dashed blue line) as a function of time, as respectively shown in (d1) for $A_{01}=1.14\pi/a$ and (d2) for $A_{02}=1.18\pi/a$  { in the logarithm scale}. The interference patterns can be successfully be explained by the analytical model of Eq.~(\ref{yield}), as shown in (f).     }
\label{fig:1}
\end{figure*}

Within the  quasi-electron description of the solid with a single-particle wavefunction $\Psi({\bf{r}},t)$,    the interacting  system of the solid with a laser pulse is  governed by the following TDSE
\begin{equation}
i \frac{ \partial \Psi({\bf{r}},t) } {\partial t} = (H_\text{free}+H_\text{int})\Psi({\bf{r}},t),
\label{eq:tdse}
\end{equation}
where $H_\text{free}$ is the field-free Hamiltonian of the quasi-electron with mass $m$, and $H_\text{int}$ is the interaction Hamiltonian between the electron and the laser field in the velocity gauge   {within} the usual dipole approximation. Taken $m= m_e$,  they are respectively given by
\begin{equation}
H_\text{free}=\frac{1}{2} {\bf{p}}^2+V({\bf{r}}),
\end{equation}
and
\begin{equation}
H_\text{int}={\bf{p}}\cdot {\bf{A}}(t), \label{Hint}
\end{equation}
in which $V({\bf{r}})$ is the effective potential satisfying the periodic condition and ${\bf{A}}(t)$ is the vector potential of the laser pulse. Since the total Hamiltonian is periodic in  space, according to the Bloch theorem, the wavefunction $\Psi({\bf{r}},t)$ can be expanded as
\begin{equation}
\Psi({\bf{r}},t) = \mathrm e^{i{\bf{q}}\cdot {\bf{r}}}u({\bf{r}},t),
\end{equation}
where the crystal momentum ${\bf{q}}$ belongs to the first Brillouin zone and $u({\bf{r}},t)$ is a periodic function which satisfies
\begin{equation}
i  \frac{\partial u({\bf{r}},t)} {\partial t}  = \left\{ \frac{[{\bf{p}}+{\bf{q}}+{\bf{A}}(t)]^2}{2}+V({\bf{r}})\right\}u({\bf{r}},t),
\end{equation}
except for an unimportant overall phase factor. One notices that, for a slowly-varying vector potential ${\bf{A}}(t)$, the total Hamiltonian adiabatically changes as the field-free Hamiltonian with a time-dependent parameter ${\bf{q}}(t) = {\bf{q}}_0+{\bf{A}}(t)$.
According to the quantum adiabatic theorem~\cite{landau2013quantum}, the population in each  eigenstate is essentially  unchanged if the relevant parameter varies slowly enough.
In other words, if the system is initially in the state $\ket{\Psi(t_0)}=\ket{n({\bf{q}}(t_0))}$, satisfying the eigenvalue equation
\begin{equation}
H_\text{free}({\bf{q}})\ket{n({\bf{q}})}=\varepsilon_n({\bf{q}})\ket{n({\bf{q}})},
\end{equation}
then we have, to the lowest order approximation, that
\begin{equation}
\ket{\Psi(t)} \approx \exp \left[-{i}\int_{t_0}^{t}d\tau\varepsilon_n({\bf{q}}(\tau))\right]\ket{n({\bf{q}}(t))},
\end{equation}
which differs from $\ket{n({\bf{q}}(t))}$ only with a  phase accumulation.
This approximation cannot be made if the energy difference of two eigenstates is not much larger than the change speed of ${\bf{q}}(t)$ (approximate to the  central frequency of the laser pulse in our case), in which case the Landau-Zener tunneling  mostly occurs.

For the demonstration of  the proposed Michelson interferometer of the Bloch electron in solids,  it is sufficient   to adopt  the simplest one-dimensional model, of which  the periodic potential is taken to be~\cite{Wu2015}
\begin{equation}
V(x) = V_0\left(1+\cos\frac{2\pi x}{a}\right),
\end{equation}
where $V_0=-0.37$ a.u. and the lattice constant $a=8$ a.u.  The first four eigenvalues are shown   in Fig.~\ref{fig:1}(a) as a function of parameter $q$, i.e, one has two valence bands~(V1, V2) and two conductance bands~(C1, C2).  The vector potential of the  pulse in Eq.~(\ref{Hint}) is given by
\begin{equation}
A(t) = A_0f(t)\cos\left(\omega t+\varphi\right),
\end{equation}
where $A_0, f(t), \omega$, and $\varphi$  is respectively the peak vector potential, envelope function, the central angular frequency, and the carrier envelope phase of the pulse. Please note that the Michelson interference can be observed in the HHG spectra induced by laser pulses with various shapes, as will be shown later.  The TDSE~(\ref{eq:tdse}) is {\it {ab initio}} solved using the   {same} numerical schemes as those detailed in our recent work~\cite{JinPRA}. For all the results presented below, we set the initial state to be at $q=0$ point of the band V2~[as marked in Fig.~\ref{fig:1}(a)], where the transition probability to the band C1 is the largest. Convergence of our results has been ensured against any change of relevant parameters in the numerical solution.

In order to clearly show interference fringes of our Michelson interferometer, we will first use a single-cycle laser pulse with a $\cos^2$ envelope $f(t)=\cos^2(\omega t/2N)$, $N=1$, $\varphi=0$, and $\omega=0.014$ a.u.~($\lambda = 3200$~nm), whose vector potential is  plotted in Fig.~\ref{fig:1}(b).  We calculate the harmonic radiation spectrum of the electron as a function of $A_0$ in a wide range of $[0, 2\pi/a]$, as shown in Fig.~\ref{fig:1}(c). In this spectra, there are a few distinct features to be noted. Apparently, at all values of $A_0$, the most probable  photon emission is  below   {the} photon energy of 0.1~a.u.,   which is induced by the intraband motion of the electron on V2~\cite{Wu2015}.  Because of the excitation to C1, there exists the harmonic emission  due to the interband transitions from C1 to V2, whose cutoff energy grows linearly with the increase of $A_0$. When  $A_0 <\pi/2a$, one can also notice a weak horizontal line of emission~(around the energy of 0.45 a.u.), which  results from the transition from V2 to V1,  similar to the atomic line transitions recently observed in the gas phase~\cite{Chang2014, Xiong2014}. Before $A_0$ reaches  the value of $\pi/a$, diagonal interference patterns are already present  and  extend all the way to  the largest  $A_0$ of $2\pi/a$, which come from the two-path contributions  for the same photon energy on the rising and falling part of the vector potential {respectively}~[cf., Fig.~\ref{fig:1}(b)].
The last and mostly important feature in Fig.~\ref{fig:1}(c) is the   distinct vertical stripes  in the top-right corner when   $A_0>\pi/a$, which is enclosed by white dotted lines.
These interference patterns originate from  the Michelson interferometry of the Bloch electron, which will be the focus of our discussions in the rest of {this Letter} .

First of all, one observes that the interference structures in this region only rely  on the value of $A_0$, independent from  the photoemission energy.
To interpret this phenomenon, we first calculate populations of C1 and C2 as a function of  time in Fig.~\ref{fig:1}(d1) for $A_{01}=1.14\pi/a$   and  in Fig.~\ref{fig:1}(d2) for  $A_{02}=1.18\pi/a$, respectively  corresponding to an interference minimal and maximum.
As can be seen, for the former case, as the vector potential first reaches $\pi/a$, a considerable population is transferred from   C1 to C2, and it will  jump  back  when the vector potential decreases down to $\pi/a$ again, which means that   a Landau-Zener tunneling happens  at the avoided crossing point `A'~[marked in Fig.~\ref{fig:1}(a)].   However, for the latter case, as shown in Fig.~\ref{fig:1}(d2), only a small amount of population  { in C2} can transfer back to C1.
Thus vertical interference patterns in the radiation spectrum are contributed by  the oscillation of population remaining in C2.
When the vector potential becomes zero, the sharp decrease of population in C2  means a  population transfer to a higher band C3~[not shown in Fig.~\ref{fig:1}(a)].

Let us first qualitatively analyze this interference process.   The  electron starts from the initial state in V2, is then excited to C1, and with the increase of the vector potential it can move towards  the avoided crossing point `A'.  It is then split into two parts, one part keeps to move along  C1 and the other moves along C2, which   corresponds to a population partition between the two bands.
Due to the energy difference between the two bands, there will be a difference of phase  accumulation when they meet again, which will   {modulate}  the population remaining in C2. In the end,  it will   {affect}  the intensity of harmonic radiation that can be measured.
The above scenario is exactly equivalent to the usual optical Michelson interferometer, as depicted in Fig.~\ref{fig:1}(e).

Within the above  picture of our Michelson interferometer in mind,  it is possible to come out with an analytical calculation of the interference patterns. For simplicity, we assume that the Landau-Zener tunneling only happens at $q=\pi/a$~[cf. Fig.~\ref{fig:1}(a)~and~(b)].   After the tunneling at the point `A', the system changes from the state $\ket{\text{C1}}$ to $\alpha_1\ket{\text{C1}}+\beta_1\ket{\text{C2}}$. After the vector potential reaches its maximum, the electron can move  back to the tunneling point    and  the system  is then in a state of   $\alpha_1\exp[-i\int_{t_i}^{t_f}\varepsilon_3(A(t))dt]\ket{\text{C1}}+\beta_1\exp[-i\int_{t_i}^{t_f}\varepsilon_4(A(t))dt]\ket{\text{C2}}$, where $t_i$ and $t_f$ are the two roots of the equation $A(t)=\pi/a$ for the single-cycle pulse.
After the second Landau-Zener tunneling, the amplitude $W$ in  C2 is
\begin{eqnarray}
W  = & & \alpha_1\beta_2\exp\left[-i\int_{t_i}^{t_f}\varepsilon_3(A(t))dt\right] \nonumber \\ &+&\alpha_2\beta_1\exp\left[-i\int_{t_i}^{t_f}\varepsilon_4(A(t))dt\right].
\end{eqnarray}
Note that  the `transmittancy' $\beta_1$ and $\beta_2$ is the Landau-Zener tunneling rate induced by the electric field,  one thus has $\beta_1=-\beta_2$ since the electric fields in the two Landau-Zener tunneling processes differ just with a  sign.
The `reflectivity' $|\alpha_1|$ is equal to $|\alpha_2|$ according to the unitarity, and the relative phase $\arg\alpha_1/\alpha_2$ can be determined by considering the tunneling process precisely~\cite{Shevchenko2010}
\begin{equation}
\arg\frac{\alpha_1}{\alpha_2} = \frac{\pi}{2}+2\delta(\ln\delta-1)+2\arg\Gamma(1-i\delta),
\end{equation}
in which $\delta=\Delta^2a/24\pi |E|$, where $\Delta$ is the band gap between the two bands and $E$ is the electric field strength at the  tunneling point.
Combining these facts, one can infer  that the resultant yield of harmonic signal $Y$ in the enclosed region  of  Fig.~\ref{fig:1}(c) satisfies
\begin{eqnarray}
 Y\propto |W|^2  \propto  1+\cos\Delta\varphi,
\label{yield}
\end{eqnarray}
with
\begin{eqnarray}
&& \Delta\varphi = \pi+\arg\frac{\alpha_1}{\alpha_2}+\int_{t_i}^{t_f}\left[\varepsilon_4(A(t))-\varepsilon_3(A(t))\right]dt.
\label{eq:phase}
\end{eqnarray}
In Fig.~\ref{fig:1}(f), we compare the analytical result of Eq.~(\ref{yield}) with that of the integrated signal calculated by TDSE.
One finds that the oscillation behaviour of  the two curves  agree with each other very well, except for the large $A_0$ region close to $2\pi/a$, where the transition to a higher band C3 is not taken into consideration in the model. Obviously, the maximum contrast ratio is observed independent of other conditions, as  in the usual Michelson interferometer in optics.

\begin{figure}
\includegraphics[width=0.95\linewidth]{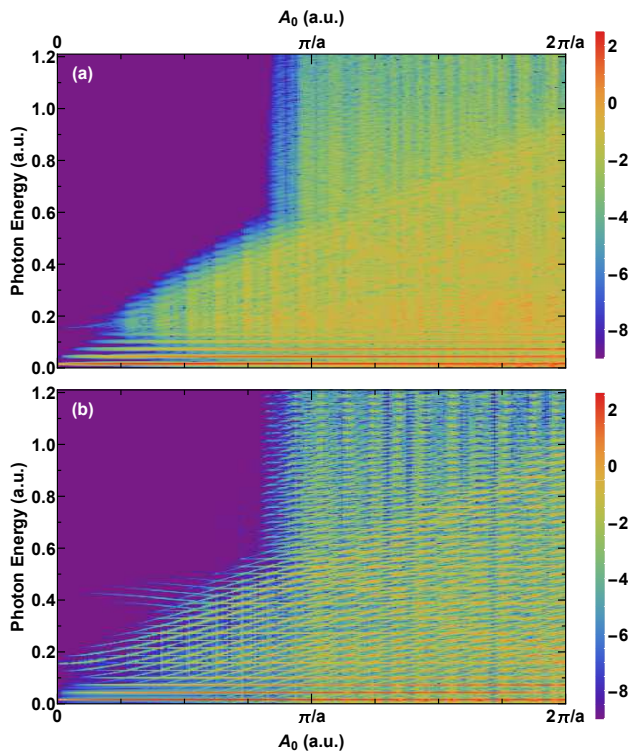}
\caption{Harmonics spectra plotted in the logarithmic scale as a function of peak vector potential $A_0$ for a longer pulse  of  an 8-cycle at the wavelength of 3200~nm for:  (a) a cosine square, and (b) a trapezoidal envelope.}
\label{fig:2}
\end{figure}

Finally, we emphasize that the Michelson interferometry in the HHG does not only exist in the single-cycle pulse.   We calculate the  HHG spectra   under the same other laser parameters by increasing the  $\cos^2$ pulse to  $N=8$, as  shown in Fig.~\ref{fig:2}(a). Vertical patterns are still clear, although much more complicated due to different maximum vector potentials in each half cycle. In addition, the vertical stripes are extended down to low-order harmonics, which is due to the fact that population distributions in a certain half-cycle have already  been affected by interferences induced by previous half-cycles.  It should be noted that this phenomena  showed up in some figures of previous theoretical studies~\cite{Higuchi2014, Wu2015, JinPRA}, but without any discussions about it. As mentioned before, the Michelson interferometry should also be present for other forms of the pulse envelope.  In Fig.~\ref{fig:2}(b),  we choose an 8-cycle laser pulse with a trapezoid envelope to guarantee the same maximum vector potentials in each half cycle.
Sharper vertical stripes are observed, whose intervals are similar to Fig.~\ref{fig:1}(c), which come from the product of multiple interferences.
 {However, it should be noted that even in this simple case, the distribution of stripes is not exactly same with that of the single-cycle case, since other types of interference channels also contribute~\cite{Shevchenko2010}.}
For $A_0<\pi/a$ , sharp parallel curves are also observed,  which may be contributed by the effects of the dressed~(Floquet) states for the multiple-cycle pulse~\cite{FloquetBloch}. In this region,  resonance-like vertical lines are also present, which are formed at the crossing points of  the dressed-state curves and the low-order harmonics. One should not mix these structures with those formed by the Michelson interferometer discussed above.

In conclusions, we have demonstrated  an interesting oscillation structure in the high-order harmonics generated in solids, which is successfully explained by  the  Michelson interferometer of the Bloch electron.  The present study deepens our understanding of the mechanism of HHG in solids and is potentially useful for the production of coherent light sources in the VUV and XUV region.   We believe that our finding can be experimentally observed  with the current technologies. One may concern about the material damage by the  applied  laser since the peak vector potential needs to be in the order of $\pi/a$. However, this problem can be avoided by using a mid-infrared laser with a wavelength up to several microns. There does  exist a  dephasing time in the solid sample, but its effects can be largely deduced by lowering the temperature.  In addition, an efficient excitation from the valence band to the conductance band can be achieved by a coherent UV sources, which will enhance the harmonic yield.  Finally, we expect that for a two-dimensional case, by adding another orthogonal pulse  after the splitting of the Bloch electron wavepacket, the Berry phase along a closed orbit in the reciprocal space may be measured. Our proposal may find potential applications in imaging the properties of  the energy bands in solids.

\begin{acknowledgments}

 This work is supported by the National Natural Science Foundation of China~(NSFC) under Grant Nos. 11725416 \& 11574010, and by the National Key R\&D Program of China~(Grant~No.~2018YFA0306302). L.Y.P. acknowledges the support by the National Science Fund for Distinguished Young Scholars.

\end{acknowledgments}

\bibliography{Ref_JPB_Letter}

\begin{thebibliography}{51}%
\makeatletter
\providecommand \@ifxundefined [1]{%
 \@ifx{#1\undefined}
}%
\providecommand \@ifnum [1]{%
 \ifnum #1\expandafter \@firstoftwo
 \else \expandafter \@secondoftwo
 \fi
}%
\providecommand \@ifx [1]{%
 \ifx #1\expandafter \@firstoftwo
 \else \expandafter \@secondoftwo
 \fi
}%
\providecommand \natexlab [1]{#1}%
\providecommand \enquote  [1]{``#1''}%
\providecommand \bibnamefont  [1]{#1}%
\providecommand \bibfnamefont [1]{#1}%
\providecommand \citenamefont [1]{#1}%
\providecommand \href@noop [0]{\@secondoftwo}%
\providecommand \href [0]{\begingroup \@sanitize@url \@href}%
\providecommand \@href[1]{\@@startlink{#1}\@@href}%
\providecommand \@@href[1]{\endgroup#1\@@endlink}%
\providecommand \@sanitize@url [0]{\catcode `\\12\catcode `\$12\catcode
  `\&12\catcode `\#12\catcode `\^12\catcode `\_12\catcode `\%12\relax}%
\providecommand \@@startlink[1]{}%
\providecommand \@@endlink[0]{}%
\providecommand \url  [0]{\begingroup\@sanitize@url \@url }%
\providecommand \@url [1]{\endgroup\@href {#1}{\urlprefix }}%
\providecommand \urlprefix  [0]{URL }%
\providecommand \Eprint [0]{\href }%
\providecommand \doibase [0]{http://dx.doi.org/}%
\providecommand \selectlanguage [0]{\@gobble}%
\providecommand \bibinfo  [0]{\@secondoftwo}%
\providecommand \bibfield  [0]{\@secondoftwo}%
\providecommand \translation [1]{[#1]}%
\providecommand \BibitemOpen [0]{}%
\providecommand \bibitemStop [0]{}%
\providecommand \bibitemNoStop [0]{.\EOS\space}%
\providecommand \EOS [0]{\spacefactor3000\relax}%
\providecommand \BibitemShut  [1]{\csname bibitem#1\endcsname}%
\let\auto@bib@innerbib\@empty
\bibitem [{\citenamefont {Brabec}\ and\ \citenamefont
  {Krausz}(2000)}]{RevModPhys.72.545}%
  \BibitemOpen
  \bibfield  {author} {\bibinfo {author} {\bibfnamefont {T.}~\bibnamefont
  {Brabec}}\ and\ \bibinfo {author} {\bibfnamefont {F.}~\bibnamefont
  {Krausz}},\ }\href {\doibase 10.1103/RevModPhys.72.545} {\bibfield  {journal}
  {\bibinfo  {journal} {Rev. Mod. Phys.}\ }\textbf {\bibinfo {volume} {72}},\
  \bibinfo {pages} {545} (\bibinfo {year} {2000})}\BibitemShut {NoStop}%
\bibitem [{\citenamefont {Hentschel}\ \emph {et~al.}(2001)\citenamefont
  {Hentschel}, \citenamefont {Kienberger}, \citenamefont {Spielmann},
  \citenamefont {Reider}, \citenamefont {Milosevic}, \citenamefont {Brabec},
  \citenamefont {Corkum}, \citenamefont {Heinzmann}, \citenamefont {Drescher},\
  and\ \citenamefont {Krausz}}]{Hentschel2001}%
  \BibitemOpen
  \bibfield  {author} {\bibinfo {author} {\bibfnamefont {M.}~\bibnamefont
  {Hentschel}}, \bibinfo {author} {\bibfnamefont {R.}~\bibnamefont
  {Kienberger}}, \bibinfo {author} {\bibfnamefont {C.}~\bibnamefont
  {Spielmann}}, \bibinfo {author} {\bibfnamefont {G.}~\bibnamefont {Reider}},
  \bibinfo {author} {\bibfnamefont {N.}~\bibnamefont {Milosevic}}, \bibinfo
  {author} {\bibfnamefont {T.}~\bibnamefont {Brabec}}, \bibinfo {author}
  {\bibfnamefont {P.}~\bibnamefont {Corkum}}, \bibinfo {author} {\bibfnamefont
  {U.}~\bibnamefont {Heinzmann}}, \bibinfo {author} {\bibfnamefont
  {M.}~\bibnamefont {Drescher}}, \ and\ \bibinfo {author} {\bibfnamefont
  {F.}~\bibnamefont {Krausz}},\ }\href {\doibase {10.1038/35107000}} {\bibfield
   {journal} {\bibinfo  {journal} {Nature}\ }\textbf {\bibinfo {volume}
  {{414}}},\ \bibinfo {pages} {{509}} (\bibinfo {year} {{2001}})}\BibitemShut
  {NoStop}%
\bibitem [{\citenamefont {Paul}\ \emph {et~al.}(2001)\citenamefont {Paul},
  \citenamefont {Toma}, \citenamefont {Breger}, \citenamefont {Mullot},
  \citenamefont {Au\'ge}, \citenamefont {Balcou}, \citenamefont {Muller},\ and\
  \citenamefont {Agostini.}}]{Paul}%
  \BibitemOpen
  \bibfield  {author} {\bibinfo {author} {\bibfnamefont {P.~M.}\ \bibnamefont
  {Paul}}, \bibinfo {author} {\bibfnamefont {E.~S.}\ \bibnamefont {Toma}},
  \bibinfo {author} {\bibfnamefont {P.}~\bibnamefont {Breger}}, \bibinfo
  {author} {\bibfnamefont {G.}~\bibnamefont {Mullot}}, \bibinfo {author}
  {\bibfnamefont {F.}~\bibnamefont {Au\'ge}}, \bibinfo {author} {\bibfnamefont
  {P.}~\bibnamefont {Balcou}}, \bibinfo {author} {\bibfnamefont {H.~G.}\
  \bibnamefont {Muller}}, \ and\ \bibinfo {author} {\bibfnamefont
  {P.}~\bibnamefont {Agostini.}},\ }\href@noop {} {\bibfield  {journal}
  {\bibinfo  {journal} {Science}\ }\textbf {\bibinfo {volume} {292}},\ \bibinfo
  {pages} {1689} (\bibinfo {year} {2001})}\BibitemShut {NoStop}%
\bibitem [{\citenamefont {Liao}\ \emph {et~al.}(2016)\citenamefont {Liao},
  \citenamefont {Li}, \citenamefont {Zhang}, \citenamefont {Liu}, \citenamefont
  {Ge}, \citenamefont {Yang}, \citenamefont {Wei}, \citenamefont {Yuan},
  \citenamefont {Deng}, \citenamefont {Zhu}, \citenamefont {Zhang},
  \citenamefont {Wang}, \citenamefont {Sheng}, \citenamefont {Chen},
  \citenamefont {Lu}, \citenamefont {Ma}, \citenamefont {Wang},\ and\
  \citenamefont {Zhang}}]{PhysRevLett.116.205003}%
  \BibitemOpen
  \bibfield  {author} {\bibinfo {author} {\bibfnamefont {G.-Q.}\ \bibnamefont
  {Liao}}, \bibinfo {author} {\bibfnamefont {Y.-T.}\ \bibnamefont {Li}},
  \bibinfo {author} {\bibfnamefont {Y.-H.}\ \bibnamefont {Zhang}}, \bibinfo
  {author} {\bibfnamefont {H.}~\bibnamefont {Liu}}, \bibinfo {author}
  {\bibfnamefont {X.-L.}\ \bibnamefont {Ge}}, \bibinfo {author} {\bibfnamefont
  {S.}~\bibnamefont {Yang}}, \bibinfo {author} {\bibfnamefont {W.-Q.}\
  \bibnamefont {Wei}}, \bibinfo {author} {\bibfnamefont {X.-H.}\ \bibnamefont
  {Yuan}}, \bibinfo {author} {\bibfnamefont {Y.-Q.}\ \bibnamefont {Deng}},
  \bibinfo {author} {\bibfnamefont {B.-J.}\ \bibnamefont {Zhu}}, \bibinfo
  {author} {\bibfnamefont {Z.}~\bibnamefont {Zhang}}, \bibinfo {author}
  {\bibfnamefont {W.-M.}\ \bibnamefont {Wang}}, \bibinfo {author}
  {\bibfnamefont {Z.-M.}\ \bibnamefont {Sheng}}, \bibinfo {author}
  {\bibfnamefont {L.-M.}\ \bibnamefont {Chen}}, \bibinfo {author}
  {\bibfnamefont {X.}~\bibnamefont {Lu}}, \bibinfo {author} {\bibfnamefont
  {J.-L.}\ \bibnamefont {Ma}}, \bibinfo {author} {\bibfnamefont
  {X.}~\bibnamefont {Wang}}, \ and\ \bibinfo {author} {\bibfnamefont
  {J.}~\bibnamefont {Zhang}},\ }\href {\doibase 10.1103/PhysRevLett.116.205003}
  {\bibfield  {journal} {\bibinfo  {journal} {Phys. Rev. Lett.}\ }\textbf
  {\bibinfo {volume} {116}},\ \bibinfo {pages} {205003} (\bibinfo {year}
  {2016})}\BibitemShut {NoStop}%
\bibitem [{\citenamefont {Kampfrath}\ \emph {et~al.}(2013)\citenamefont
  {Kampfrath}, \citenamefont {Tanaka},\ and\ \citenamefont
  {Nelson}}]{Kampfrath2013}%
  \BibitemOpen
  \bibfield  {author} {\bibinfo {author} {\bibfnamefont {T.}~\bibnamefont
  {Kampfrath}}, \bibinfo {author} {\bibfnamefont {K.}~\bibnamefont {Tanaka}}, \
  and\ \bibinfo {author} {\bibfnamefont {K.~A.}\ \bibnamefont {Nelson}},\
  }\href {\doibase 10.1038/nphoton.2013.184} {\bibfield  {journal} {\bibinfo
  {journal} {Nat. Photonics}\ }\textbf {\bibinfo {volume} {7}},\ \bibinfo
  {pages} {680} (\bibinfo {year} {2013})}\BibitemShut {NoStop}%
\bibitem [{\citenamefont {Corkum}\ and\ \citenamefont
  {Krausz}(2007)}]{Corkum2007}%
  \BibitemOpen
  \bibfield  {author} {\bibinfo {author} {\bibfnamefont {P.~B.}\ \bibnamefont
  {Corkum}}\ and\ \bibinfo {author} {\bibfnamefont {F.}~\bibnamefont
  {Krausz}},\ }\href {\doibase {10.1038/nphys620}} {\bibfield  {journal}
  {\bibinfo  {journal} {Nat. Phys.}\ }\textbf {\bibinfo {volume} {{3}}},\
  \bibinfo {pages} {{381}} (\bibinfo {year} {{2007}})}\BibitemShut {NoStop}%
\bibitem [{\citenamefont {Krausz}\ and\ \citenamefont {Ivanov}(2009)}]{Krausz}%
  \BibitemOpen
  \bibfield  {author} {\bibinfo {author} {\bibfnamefont {F.}~\bibnamefont
  {Krausz}}\ and\ \bibinfo {author} {\bibfnamefont {M.}~\bibnamefont
  {Ivanov}},\ }\href@noop {} {\bibfield  {journal} {\bibinfo  {journal} {Rev.
  Mod. Phys.}\ }\textbf {\bibinfo {volume} {81}},\ \bibinfo {pages} {163}
  (\bibinfo {year} {2009})}\BibitemShut {NoStop}%
\bibitem [{\citenamefont {Peng}\ \emph {et~al.}(2015)\citenamefont {Peng},
  \citenamefont {Jiang}, \citenamefont {Geng}, \citenamefont {Xiong},\ and\
  \citenamefont {Gong}}]{PENG20151}%
  \BibitemOpen
  \bibfield  {author} {\bibinfo {author} {\bibfnamefont {L.-Y.}\ \bibnamefont
  {Peng}}, \bibinfo {author} {\bibfnamefont {W.-C.}\ \bibnamefont {Jiang}},
  \bibinfo {author} {\bibfnamefont {J.-W.}\ \bibnamefont {Geng}}, \bibinfo
  {author} {\bibfnamefont {W.-H.}\ \bibnamefont {Xiong}}, \ and\ \bibinfo
  {author} {\bibfnamefont {Q.}~\bibnamefont {Gong}},\ }\href {\doibase
  https://doi.org/10.1016/j.physrep.2015.02.002} {\bibfield  {journal}
  {\bibinfo  {journal} {Phys. Rep.}\ }\textbf {\bibinfo {volume} {575}},\
  \bibinfo {pages} {1 } (\bibinfo {year} {2015})}\BibitemShut {NoStop}%
\bibitem [{\citenamefont {Schubert}\ \emph {et~al.}(2014)\citenamefont
  {Schubert}, \citenamefont {Hohenleutner}, \citenamefont {Langer},
  \citenamefont {Urbanek}, \citenamefont {Lange}, \citenamefont {Huttner},
  \citenamefont {Golde}, \citenamefont {Meier}, \citenamefont {Kira},
  \citenamefont {Koch},\ and\ \citenamefont {Huber}}]{Schubert2014}%
  \BibitemOpen
  \bibfield  {author} {\bibinfo {author} {\bibfnamefont {O.}~\bibnamefont
  {Schubert}}, \bibinfo {author} {\bibfnamefont {M.}~\bibnamefont
  {Hohenleutner}}, \bibinfo {author} {\bibfnamefont {F.}~\bibnamefont
  {Langer}}, \bibinfo {author} {\bibfnamefont {B.}~\bibnamefont {Urbanek}},
  \bibinfo {author} {\bibfnamefont {C.}~\bibnamefont {Lange}}, \bibinfo
  {author} {\bibfnamefont {U.}~\bibnamefont {Huttner}}, \bibinfo {author}
  {\bibfnamefont {D.}~\bibnamefont {Golde}}, \bibinfo {author} {\bibfnamefont
  {T.}~\bibnamefont {Meier}}, \bibinfo {author} {\bibfnamefont
  {M.}~\bibnamefont {Kira}}, \bibinfo {author} {\bibfnamefont {S.~W.}\
  \bibnamefont {Koch}}, \ and\ \bibinfo {author} {\bibfnamefont
  {R.}~\bibnamefont {Huber}},\ }\href {\doibase 10.1038/nphoton.2013.349}
  {\bibfield  {journal} {\bibinfo  {journal} {Nat. Photonics}\ }\textbf
  {\bibinfo {volume} {8}},\ \bibinfo {pages} {119} (\bibinfo {year}
  {2014})}\BibitemShut {NoStop}%
\bibitem [{\citenamefont {Schultze}\ \emph {et~al.}(2014)\citenamefont
  {Schultze}, \citenamefont {Ramasesha}, \citenamefont {Pemmaraju},
  \citenamefont {Sato}, \citenamefont {Whitmore}, \citenamefont {Gandman},
  \citenamefont {Prell}, \citenamefont {Borja}, \citenamefont {Prendergast},
  \citenamefont {Yabana}, \citenamefont {Neumark},\ and\ \citenamefont
  {Leone}}]{Schultze2014}%
  \BibitemOpen
  \bibfield  {author} {\bibinfo {author} {\bibfnamefont {M.}~\bibnamefont
  {Schultze}}, \bibinfo {author} {\bibfnamefont {K.}~\bibnamefont {Ramasesha}},
  \bibinfo {author} {\bibfnamefont {C.~D.}\ \bibnamefont {Pemmaraju}}, \bibinfo
  {author} {\bibfnamefont {S.~A.}\ \bibnamefont {Sato}}, \bibinfo {author}
  {\bibfnamefont {D.}~\bibnamefont {Whitmore}}, \bibinfo {author}
  {\bibfnamefont {A.}~\bibnamefont {Gandman}}, \bibinfo {author} {\bibfnamefont
  {J.~S.}\ \bibnamefont {Prell}}, \bibinfo {author} {\bibfnamefont {L.~J.}\
  \bibnamefont {Borja}}, \bibinfo {author} {\bibfnamefont {D.}~\bibnamefont
  {Prendergast}}, \bibinfo {author} {\bibfnamefont {K.}~\bibnamefont {Yabana}},
  \bibinfo {author} {\bibfnamefont {D.~M.}\ \bibnamefont {Neumark}}, \ and\
  \bibinfo {author} {\bibfnamefont {S.~R.}\ \bibnamefont {Leone}},\ }\href
  {\doibase 10.1126/science.1260311} {\bibfield  {journal} {\bibinfo  {journal}
  {Science}\ }\textbf {\bibinfo {volume} {346}},\ \bibinfo {pages} {1348}
  (\bibinfo {year} {2014})}\BibitemShut {NoStop}%
\bibitem [{\citenamefont {Higuchi}\ \emph {et~al.}(2017)\citenamefont
  {Higuchi}, \citenamefont {Heide}, \citenamefont {Ullmann}, \citenamefont
  {Weber},\ and\ \citenamefont {Hommelhoff}}]{Higuchi2017a}%
  \BibitemOpen
  \bibfield  {author} {\bibinfo {author} {\bibfnamefont {T.}~\bibnamefont
  {Higuchi}}, \bibinfo {author} {\bibfnamefont {C.}~\bibnamefont {Heide}},
  \bibinfo {author} {\bibfnamefont {K.}~\bibnamefont {Ullmann}}, \bibinfo
  {author} {\bibfnamefont {H.~B.}\ \bibnamefont {Weber}}, \ and\ \bibinfo
  {author} {\bibfnamefont {P.}~\bibnamefont {Hommelhoff}},\ }\href {\doibase
  10.1038/nature23900} {\bibfield  {journal} {\bibinfo  {journal} {Nature}\
  }\textbf {\bibinfo {volume} {550}},\ \bibinfo {pages} {224} (\bibinfo {year}
  {2017})}\BibitemShut {NoStop}%
\bibitem [{\citenamefont {McPherson}\ \emph {et~al.}(1987)\citenamefont
  {McPherson}, \citenamefont {Gibson}, \citenamefont {Jara}, \citenamefont
  {Johann}, \citenamefont {Luk}, \citenamefont {McIntyre}, \citenamefont
  {Boyer},\ and\ \citenamefont {Rhodes}}]{McPherson1987}%
  \BibitemOpen
  \bibfield  {author} {\bibinfo {author} {\bibfnamefont {A.}~\bibnamefont
  {McPherson}}, \bibinfo {author} {\bibfnamefont {G.}~\bibnamefont {Gibson}},
  \bibinfo {author} {\bibfnamefont {H.}~\bibnamefont {Jara}}, \bibinfo {author}
  {\bibfnamefont {U.}~\bibnamefont {Johann}}, \bibinfo {author} {\bibfnamefont
  {T.~S.}\ \bibnamefont {Luk}}, \bibinfo {author} {\bibfnamefont {I.~A.}\
  \bibnamefont {McIntyre}}, \bibinfo {author} {\bibfnamefont {K.}~\bibnamefont
  {Boyer}}, \ and\ \bibinfo {author} {\bibfnamefont {C.~K.}\ \bibnamefont
  {Rhodes}},\ }\href {\doibase 10.1364/JOSAB.4.000595} {\bibfield  {journal}
  {\bibinfo  {journal} {J. Opt. Soc. Am. B}\ }\textbf {\bibinfo {volume} {4}},\
  \bibinfo {pages} {595} (\bibinfo {year} {1987})}\BibitemShut {NoStop}%
\bibitem [{\citenamefont {Ferray}\ \emph {et~al.}(1988)\citenamefont {Ferray},
  \citenamefont {L'Huillier}, \citenamefont {Li}, \citenamefont {Lompre},
  \citenamefont {Mainfray},\ and\ \citenamefont {Manus}}]{1-3-FERRAY1988}%
  \BibitemOpen
  \bibfield  {author} {\bibinfo {author} {\bibfnamefont {M.}~\bibnamefont
  {Ferray}}, \bibinfo {author} {\bibfnamefont {A.}~\bibnamefont {L'Huillier}},
  \bibinfo {author} {\bibfnamefont {X.}~\bibnamefont {Li}}, \bibinfo {author}
  {\bibfnamefont {L.~A.}\ \bibnamefont {Lompre}}, \bibinfo {author}
  {\bibfnamefont {G.}~\bibnamefont {Mainfray}}, \ and\ \bibinfo {author}
  {\bibfnamefont {C.}~\bibnamefont {Manus}},\ }\href {\doibase
  {10.1088/0953-4075/21/3/001}} {\bibfield  {journal} {\bibinfo  {journal} {J.
  Phys. B}\ }\textbf {\bibinfo {volume} {{21}}},\ \bibinfo {pages} {{L31}}
  (\bibinfo {year} {{1988}})}\BibitemShut {NoStop}%
\bibitem [{\citenamefont {Ghimire}\ \emph {et~al.}(2011)\citenamefont
  {Ghimire}, \citenamefont {Dichiara}, \citenamefont {Sistrunk}, \citenamefont
  {Agostini}, \citenamefont {Dimauro},\ and\ \citenamefont
  {Reis}}]{Ghimire2011}%
  \BibitemOpen
  \bibfield  {author} {\bibinfo {author} {\bibfnamefont {S.}~\bibnamefont
  {Ghimire}}, \bibinfo {author} {\bibfnamefont {A.~D.}\ \bibnamefont
  {Dichiara}}, \bibinfo {author} {\bibfnamefont {E.}~\bibnamefont {Sistrunk}},
  \bibinfo {author} {\bibfnamefont {P.}~\bibnamefont {Agostini}}, \bibinfo
  {author} {\bibfnamefont {L.~F.}\ \bibnamefont {Dimauro}}, \ and\ \bibinfo
  {author} {\bibfnamefont {D.~A.}\ \bibnamefont {Reis}},\ }\href {\doibase
  10.1038/nphys1847} {\bibfield  {journal} {\bibinfo  {journal} {Nat. Phys.}\
  }\textbf {\bibinfo {volume} {7}},\ \bibinfo {pages} {138} (\bibinfo {year}
  {2011})}\BibitemShut {NoStop}%
\bibitem [{\citenamefont {Vampa}\ \emph
  {et~al.}(2015{\natexlab{a}})\citenamefont {Vampa}, \citenamefont {Hammond},
  \citenamefont {Thir{\'{e}}}, \citenamefont {Schmidt}, \citenamefont
  {L{\'{e}}gar{\'{e}}}, \citenamefont {McDonald}, \citenamefont {Brabec},\ and\
  \citenamefont {Corkum}}]{Vampa2015b}%
  \BibitemOpen
  \bibfield  {author} {\bibinfo {author} {\bibfnamefont {G.}~\bibnamefont
  {Vampa}}, \bibinfo {author} {\bibfnamefont {T.~J.}\ \bibnamefont {Hammond}},
  \bibinfo {author} {\bibfnamefont {N.}~\bibnamefont {Thir{\'{e}}}}, \bibinfo
  {author} {\bibfnamefont {B.~E.}\ \bibnamefont {Schmidt}}, \bibinfo {author}
  {\bibfnamefont {F.}~\bibnamefont {L{\'{e}}gar{\'{e}}}}, \bibinfo {author}
  {\bibfnamefont {C.~R.}\ \bibnamefont {McDonald}}, \bibinfo {author}
  {\bibfnamefont {T.}~\bibnamefont {Brabec}}, \ and\ \bibinfo {author}
  {\bibfnamefont {P.~B.}\ \bibnamefont {Corkum}},\ }\href {\doibase
  10.1038/nature14517} {\bibfield  {journal} {\bibinfo  {journal} {Nature}\
  }\textbf {\bibinfo {volume} {522}},\ \bibinfo {pages} {462} (\bibinfo {year}
  {2015}{\natexlab{a}})}\BibitemShut {NoStop}%
\bibitem [{\citenamefont {Luu}\ \emph {et~al.}(2015)\citenamefont {Luu},
  \citenamefont {Garg}, \citenamefont {{Yu. Kruchinin}}, \citenamefont
  {Moulet}, \citenamefont {Hassan},\ and\ \citenamefont
  {Goulielmakis}}]{Luu2015}%
  \BibitemOpen
  \bibfield  {author} {\bibinfo {author} {\bibfnamefont {T.~T.}\ \bibnamefont
  {Luu}}, \bibinfo {author} {\bibfnamefont {M.}~\bibnamefont {Garg}}, \bibinfo
  {author} {\bibfnamefont {S.}~\bibnamefont {{Yu. Kruchinin}}}, \bibinfo
  {author} {\bibfnamefont {A.}~\bibnamefont {Moulet}}, \bibinfo {author}
  {\bibfnamefont {M.~T.}\ \bibnamefont {Hassan}}, \ and\ \bibinfo {author}
  {\bibfnamefont {E.}~\bibnamefont {Goulielmakis}},\ }\href {\doibase
  10.1038/nature14456} {\bibfield  {journal} {\bibinfo  {journal} {Nature}\
  }\textbf {\bibinfo {volume} {521}},\ \bibinfo {pages} {498} (\bibinfo {year}
  {2015})}\BibitemShut {NoStop}%
\bibitem [{\citenamefont {Ndabashimiye}\ \emph {et~al.}(2016)\citenamefont
  {Ndabashimiye}, \citenamefont {Ghimire}, \citenamefont {Wu}, \citenamefont
  {Browne}, \citenamefont {Schafer}, \citenamefont {Gaarde},\ and\
  \citenamefont {Reis}}]{Ndabashimiye2016}%
  \BibitemOpen
  \bibfield  {author} {\bibinfo {author} {\bibfnamefont {G.}~\bibnamefont
  {Ndabashimiye}}, \bibinfo {author} {\bibfnamefont {S.}~\bibnamefont
  {Ghimire}}, \bibinfo {author} {\bibfnamefont {M.}~\bibnamefont {Wu}},
  \bibinfo {author} {\bibfnamefont {D.~A.}\ \bibnamefont {Browne}}, \bibinfo
  {author} {\bibfnamefont {K.~J.}\ \bibnamefont {Schafer}}, \bibinfo {author}
  {\bibfnamefont {M.~B.}\ \bibnamefont {Gaarde}}, \ and\ \bibinfo {author}
  {\bibfnamefont {D.~A.}\ \bibnamefont {Reis}},\ }\href {\doibase
  10.1038/nature17660} {\bibfield  {journal} {\bibinfo  {journal} {Nature}\
  }\textbf {\bibinfo {volume} {534}},\ \bibinfo {pages} {520} (\bibinfo {year}
  {2016})}\BibitemShut {NoStop}%
\bibitem [{\citenamefont {You}\ \emph {et~al.}(2017)\citenamefont {You},
  \citenamefont {Reis},\ and\ \citenamefont {Ghimire}}]{You2017}%
  \BibitemOpen
  \bibfield  {author} {\bibinfo {author} {\bibfnamefont {Y.~S.}\ \bibnamefont
  {You}}, \bibinfo {author} {\bibfnamefont {D.~A.}\ \bibnamefont {Reis}}, \
  and\ \bibinfo {author} {\bibfnamefont {S.}~\bibnamefont {Ghimire}},\ }\href
  {\doibase 10.1038/nphys3955} {\bibfield  {journal} {\bibinfo  {journal} {Nat.
  Phys.}\ }\textbf {\bibinfo {volume} {13}},\ \bibinfo {pages} {345} (\bibinfo
  {year} {2017})}\BibitemShut {NoStop}%
\bibitem [{\citenamefont {Sivis}\ \emph {et~al.}(2017)\citenamefont {Sivis},
  \citenamefont {Taucer}, \citenamefont {Vampa}, \citenamefont {Johnston},
  \citenamefont {Staudte}, \citenamefont {Naumov}, \citenamefont {Villeneuve},
  \citenamefont {Ropers},\ and\ \citenamefont {Corkum}}]{Sivis00041}%
  \BibitemOpen
  \bibfield  {author} {\bibinfo {author} {\bibfnamefont {M.}~\bibnamefont
  {Sivis}}, \bibinfo {author} {\bibfnamefont {M.}~\bibnamefont {Taucer}},
  \bibinfo {author} {\bibfnamefont {G.}~\bibnamefont {Vampa}}, \bibinfo
  {author} {\bibfnamefont {K.}~\bibnamefont {Johnston}}, \bibinfo {author}
  {\bibfnamefont {A.}~\bibnamefont {Staudte}}, \bibinfo {author} {\bibfnamefont
  {A.~Y.}\ \bibnamefont {Naumov}}, \bibinfo {author} {\bibfnamefont {D.~M.}\
  \bibnamefont {Villeneuve}}, \bibinfo {author} {\bibfnamefont
  {C.}~\bibnamefont {Ropers}}, \ and\ \bibinfo {author} {\bibfnamefont {P.~B.}\
  \bibnamefont {Corkum}},\ }\href@noop {} {\bibfield  {journal} {\bibinfo
  {journal} {{Science}}\ }\textbf {\bibinfo {volume} {{357}}},\ \bibinfo
  {pages} {{303}} (\bibinfo {year} {{2017}})}\BibitemShut {NoStop}%
\bibitem [{\citenamefont {Yoshikawa}\ \emph {et~al.}(2017)\citenamefont
  {Yoshikawa}, \citenamefont {Tamaya},\ and\ \citenamefont
  {Tanaka}}]{Yoshikawa2017}%
  \BibitemOpen
  \bibfield  {author} {\bibinfo {author} {\bibfnamefont {N.}~\bibnamefont
  {Yoshikawa}}, \bibinfo {author} {\bibfnamefont {T.}~\bibnamefont {Tamaya}}, \
  and\ \bibinfo {author} {\bibfnamefont {K.}~\bibnamefont {Tanaka}},\ }\href
  {\doibase 10.1126/science.aam8861} {\bibfield  {journal} {\bibinfo  {journal}
  {Science}\ }\textbf {\bibinfo {volume} {356}},\ \bibinfo {pages} {736}
  (\bibinfo {year} {2017})}\BibitemShut {NoStop}%
\bibitem [{\citenamefont {Tamaya}\ \emph {et~al.}(2016)\citenamefont {Tamaya},
  \citenamefont {Ishikawa}, \citenamefont {Ogawa},\ and\ \citenamefont
  {Tanaka}}]{Tamaya2016}%
  \BibitemOpen
  \bibfield  {author} {\bibinfo {author} {\bibfnamefont {T.}~\bibnamefont
  {Tamaya}}, \bibinfo {author} {\bibfnamefont {A.}~\bibnamefont {Ishikawa}},
  \bibinfo {author} {\bibfnamefont {T.}~\bibnamefont {Ogawa}}, \ and\ \bibinfo
  {author} {\bibfnamefont {K.}~\bibnamefont {Tanaka}},\ }\href {\doibase
  10.1103/PhysRevLett.116.016601} {\bibfield  {journal} {\bibinfo  {journal}
  {Phys. Rev. Lett.}\ }\textbf {\bibinfo {volume} {116}},\ \bibinfo {pages}
  {016601} (\bibinfo {year} {2016})}\BibitemShut {NoStop}%
\bibitem [{\citenamefont {Jiang}\ \emph {et~al.}(2018)\citenamefont {Jiang},
  \citenamefont {Chen}, \citenamefont {Wei}, \citenamefont {Yu}, \citenamefont
  {Lu},\ and\ \citenamefont {Lin}}]{JiangLu2018}%
  \BibitemOpen
  \bibfield  {author} {\bibinfo {author} {\bibfnamefont {S.}~\bibnamefont
  {Jiang}}, \bibinfo {author} {\bibfnamefont {J.}~\bibnamefont {Chen}},
  \bibinfo {author} {\bibfnamefont {H.}~\bibnamefont {Wei}}, \bibinfo {author}
  {\bibfnamefont {C.}~\bibnamefont {Yu}}, \bibinfo {author} {\bibfnamefont
  {R.}~\bibnamefont {Lu}}, \ and\ \bibinfo {author} {\bibfnamefont {C.~D.}\
  \bibnamefont {Lin}},\ }\href {\doibase 10.1103/PhysRevLett.120.253201}
  {\bibfield  {journal} {\bibinfo  {journal} {Phys. Rev. Lett.}\ }\textbf
  {\bibinfo {volume} {120}},\ \bibinfo {pages} {253201} (\bibinfo {year}
  {2018})}\BibitemShut {NoStop}%
\bibitem [{\citenamefont {Schiffrin}\ \emph {et~al.}(2013)\citenamefont
  {Schiffrin}, \citenamefont {Paasch-Colberg}, \citenamefont {Karpowicz},
  \citenamefont {Apalkov}, \citenamefont {Gerster}, \citenamefont
  {Mühlbrandt}, \citenamefont {Korbman}, \citenamefont {Reichert},
  \citenamefont {Schultze}, \citenamefont {Holzner}, \citenamefont {Barth},
  \citenamefont {Kienberger}, \citenamefont {Ernstorfer}, \citenamefont
  {Yakovlev}, \citenamefont {Stockman},\ and\ \citenamefont
  {Krausz}}]{schiffrin_optical-field-induced_2013}%
  \BibitemOpen
  \bibfield  {author} {\bibinfo {author} {\bibfnamefont {A.}~\bibnamefont
  {Schiffrin}}, \bibinfo {author} {\bibfnamefont {T.}~\bibnamefont
  {Paasch-Colberg}}, \bibinfo {author} {\bibfnamefont {N.}~\bibnamefont
  {Karpowicz}}, \bibinfo {author} {\bibfnamefont {V.}~\bibnamefont {Apalkov}},
  \bibinfo {author} {\bibfnamefont {D.}~\bibnamefont {Gerster}}, \bibinfo
  {author} {\bibfnamefont {S.}~\bibnamefont {Mühlbrandt}}, \bibinfo {author}
  {\bibfnamefont {M.}~\bibnamefont {Korbman}}, \bibinfo {author} {\bibfnamefont
  {J.}~\bibnamefont {Reichert}}, \bibinfo {author} {\bibfnamefont
  {M.}~\bibnamefont {Schultze}}, \bibinfo {author} {\bibfnamefont
  {S.}~\bibnamefont {Holzner}}, \bibinfo {author} {\bibfnamefont {J.~V.}\
  \bibnamefont {Barth}}, \bibinfo {author} {\bibfnamefont {R.}~\bibnamefont
  {Kienberger}}, \bibinfo {author} {\bibfnamefont {R.}~\bibnamefont
  {Ernstorfer}}, \bibinfo {author} {\bibfnamefont {V.~S.}\ \bibnamefont
  {Yakovlev}}, \bibinfo {author} {\bibfnamefont {M.~I.}\ \bibnamefont
  {Stockman}}, \ and\ \bibinfo {author} {\bibfnamefont {F.}~\bibnamefont
  {Krausz}},\ }\href {\doibase 10.1038/nature11567} {\bibfield  {journal}
  {\bibinfo  {journal} {Nature}\ }\textbf {\bibinfo {volume} {493}},\ \bibinfo
  {pages} {70} (\bibinfo {year} {2013})}\BibitemShut {NoStop}%
\bibitem [{\citenamefont {Schultze}\ \emph {et~al.}(2013)\citenamefont
  {Schultze}, \citenamefont {Bothschafter}, \citenamefont {Sommer},
  \citenamefont {Holzner}, \citenamefont {Schweinberger}, \citenamefont
  {Fiess}, \citenamefont {Hofstetter}, \citenamefont {Kienberger},
  \citenamefont {Apalkov}, \citenamefont {Yakovlev}, \citenamefont {Stockman},\
  and\ \citenamefont {Krausz}}]{Schultze2013}%
  \BibitemOpen
  \bibfield  {author} {\bibinfo {author} {\bibfnamefont {M.}~\bibnamefont
  {Schultze}}, \bibinfo {author} {\bibfnamefont {E.~M.}\ \bibnamefont
  {Bothschafter}}, \bibinfo {author} {\bibfnamefont {A.}~\bibnamefont
  {Sommer}}, \bibinfo {author} {\bibfnamefont {S.}~\bibnamefont {Holzner}},
  \bibinfo {author} {\bibfnamefont {W.}~\bibnamefont {Schweinberger}}, \bibinfo
  {author} {\bibfnamefont {M.}~\bibnamefont {Fiess}}, \bibinfo {author}
  {\bibfnamefont {M.}~\bibnamefont {Hofstetter}}, \bibinfo {author}
  {\bibfnamefont {R.}~\bibnamefont {Kienberger}}, \bibinfo {author}
  {\bibfnamefont {V.}~\bibnamefont {Apalkov}}, \bibinfo {author} {\bibfnamefont
  {V.~S.}\ \bibnamefont {Yakovlev}}, \bibinfo {author} {\bibfnamefont {M.~I.}\
  \bibnamefont {Stockman}}, \ and\ \bibinfo {author} {\bibfnamefont
  {F.}~\bibnamefont {Krausz}},\ }\href {\doibase 10.1038/nature11720}
  {\bibfield  {journal} {\bibinfo  {journal} {Nature}\ }\textbf {\bibinfo
  {volume} {493}},\ \bibinfo {pages} {75} (\bibinfo {year} {2013})}\BibitemShut
  {NoStop}%
\bibitem [{\citenamefont {Vampa}\ \emph
  {et~al.}(2015{\natexlab{b}})\citenamefont {Vampa}, \citenamefont {Hammond},
  \citenamefont {Thir{\'{e}}}, \citenamefont {Schmidt}, \citenamefont
  {L{\'{e}}gar{\'{e}}}, \citenamefont {McDonald}, \citenamefont {Brabec},
  \citenamefont {Klug},\ and\ \citenamefont {Corkum}}]{Vampa2015a}%
  \BibitemOpen
  \bibfield  {author} {\bibinfo {author} {\bibfnamefont {G.}~\bibnamefont
  {Vampa}}, \bibinfo {author} {\bibfnamefont {T.~J.}\ \bibnamefont {Hammond}},
  \bibinfo {author} {\bibfnamefont {N.}~\bibnamefont {Thir{\'{e}}}}, \bibinfo
  {author} {\bibfnamefont {B.~E.}\ \bibnamefont {Schmidt}}, \bibinfo {author}
  {\bibfnamefont {F.}~\bibnamefont {L{\'{e}}gar{\'{e}}}}, \bibinfo {author}
  {\bibfnamefont {C.~R.}\ \bibnamefont {McDonald}}, \bibinfo {author}
  {\bibfnamefont {T.}~\bibnamefont {Brabec}}, \bibinfo {author} {\bibfnamefont
  {D.~D.}\ \bibnamefont {Klug}}, \ and\ \bibinfo {author} {\bibfnamefont
  {P.~B.}\ \bibnamefont {Corkum}},\ }\href {\doibase
  10.1103/PhysRevLett.115.193603} {\bibfield  {journal} {\bibinfo  {journal}
  {Phys. Rev. Lett.}\ }\textbf {\bibinfo {volume} {115}},\ \bibinfo {pages}
  {193603} (\bibinfo {year} {2015}{\natexlab{b}})}\BibitemShut {NoStop}%
\bibitem [{\citenamefont {Garg}\ \emph {et~al.}(2016)\citenamefont {Garg},
  \citenamefont {Zhan}, \citenamefont {Luu}, \citenamefont {Lakhotia},
  \citenamefont {Klostermann}, \citenamefont {Guggenmos},\ and\ \citenamefont
  {Goulielmakis}}]{Garg2016}%
  \BibitemOpen
  \bibfield  {author} {\bibinfo {author} {\bibfnamefont {M.}~\bibnamefont
  {Garg}}, \bibinfo {author} {\bibfnamefont {M.}~\bibnamefont {Zhan}}, \bibinfo
  {author} {\bibfnamefont {T.~T.}\ \bibnamefont {Luu}}, \bibinfo {author}
  {\bibfnamefont {H.}~\bibnamefont {Lakhotia}}, \bibinfo {author}
  {\bibfnamefont {T.}~\bibnamefont {Klostermann}}, \bibinfo {author}
  {\bibfnamefont {A.}~\bibnamefont {Guggenmos}}, \ and\ \bibinfo {author}
  {\bibfnamefont {E.}~\bibnamefont {Goulielmakis}},\ }\href {\doibase
  10.1038/nature19821} {\bibfield  {journal} {\bibinfo  {journal} {Nature}\
  }\textbf {\bibinfo {volume} {538}},\ \bibinfo {pages} {359} (\bibinfo {year}
  {2016})}\BibitemShut {NoStop}%
\bibitem [{\citenamefont {Luu}\ and\ \citenamefont
  {Wörner}(2018)}]{luu_measurement_2018}%
  \BibitemOpen
  \bibfield  {author} {\bibinfo {author} {\bibfnamefont {T.~T.}\ \bibnamefont
  {Luu}}\ and\ \bibinfo {author} {\bibfnamefont {H.~J.}\ \bibnamefont
  {Wörner}},\ }\href {\doibase 10.1038/s41467-018-03397-4} {\bibfield
  {journal} {\bibinfo  {journal} {Nat. Commun.}\ }\textbf {\bibinfo {volume}
  {9}},\ \bibinfo {pages} {916} (\bibinfo {year} {2018})}\BibitemShut {NoStop}%
\bibitem [{\citenamefont {Michelson}\ and\ \citenamefont
  {Morley}(1887)}]{Michelson1887}%
  \BibitemOpen
  \bibfield  {author} {\bibinfo {author} {\bibfnamefont {A.}~\bibnamefont
  {Michelson}}\ and\ \bibinfo {author} {\bibfnamefont {E.}~\bibnamefont
  {Morley}},\ }\href {\doibase 10.2475/ajs.s3-34.203.333} {\bibfield  {journal}
  {\bibinfo  {journal} {Am. J. Sci.}\ }\textbf {\bibinfo {volume} {34}},\
  \bibinfo {pages} {333} (\bibinfo {year} {1887})}\BibitemShut {NoStop}%
\bibitem [{\citenamefont {Santori}\ \emph {et~al.}(2002)\citenamefont
  {Santori}, \citenamefont {Fattal}, \citenamefont {Vu{\v{c}}kovi{\'{c}}},
  \citenamefont {Solomon},\ and\ \citenamefont {Yamamoto}}]{Santori2002}%
  \BibitemOpen
  \bibfield  {author} {\bibinfo {author} {\bibfnamefont {C.}~\bibnamefont
  {Santori}}, \bibinfo {author} {\bibfnamefont {D.}~\bibnamefont {Fattal}},
  \bibinfo {author} {\bibfnamefont {J.}~\bibnamefont {Vu{\v{c}}kovi{\'{c}}}},
  \bibinfo {author} {\bibfnamefont {G.~S.}\ \bibnamefont {Solomon}}, \ and\
  \bibinfo {author} {\bibfnamefont {Y.}~\bibnamefont {Yamamoto}},\ }\href
  {\doibase 10.1038/nature01086} {\bibfield  {journal} {\bibinfo  {journal}
  {Nature}\ }\textbf {\bibinfo {volume} {419}},\ \bibinfo {pages} {594}
  (\bibinfo {year} {2002})}\BibitemShut {NoStop}%
\bibitem [{\citenamefont {Dolling}\ \emph {et~al.}(2006)\citenamefont
  {Dolling}, \citenamefont {Enkrich}, \citenamefont {Wegener}, \citenamefont
  {Soukoulis},\ and\ \citenamefont {Linden}}]{Dolling2006}%
  \BibitemOpen
  \bibfield  {author} {\bibinfo {author} {\bibfnamefont {G.}~\bibnamefont
  {Dolling}}, \bibinfo {author} {\bibfnamefont {C.}~\bibnamefont {Enkrich}},
  \bibinfo {author} {\bibfnamefont {M.}~\bibnamefont {Wegener}}, \bibinfo
  {author} {\bibfnamefont {C.~M.}\ \bibnamefont {Soukoulis}}, \ and\ \bibinfo
  {author} {\bibfnamefont {S.}~\bibnamefont {Linden}},\ }\href {\doibase
  10.1126/science.1126021} {\bibfield  {journal} {\bibinfo  {journal}
  {Science}\ }\textbf {\bibinfo {volume} {312}},\ \bibinfo {pages} {892}
  (\bibinfo {year} {2006})}\BibitemShut {NoStop}%
\bibitem [{\citenamefont {Usenko}\ \emph {et~al.}(2017)\citenamefont {Usenko},
  \citenamefont {Przystawik}, \citenamefont {Jakob}, \citenamefont {Lazzarino},
  \citenamefont {Brenner}, \citenamefont {Toleikis}, \citenamefont {Haunhorst},
  \citenamefont {Kip},\ and\ \citenamefont {Laarmann}}]{Usenko2017}%
  \BibitemOpen
  \bibfield  {author} {\bibinfo {author} {\bibfnamefont {S.}~\bibnamefont
  {Usenko}}, \bibinfo {author} {\bibfnamefont {A.}~\bibnamefont {Przystawik}},
  \bibinfo {author} {\bibfnamefont {M.~A.}\ \bibnamefont {Jakob}}, \bibinfo
  {author} {\bibfnamefont {L.~L.}\ \bibnamefont {Lazzarino}}, \bibinfo {author}
  {\bibfnamefont {G.}~\bibnamefont {Brenner}}, \bibinfo {author} {\bibfnamefont
  {S.}~\bibnamefont {Toleikis}}, \bibinfo {author} {\bibfnamefont
  {C.}~\bibnamefont {Haunhorst}}, \bibinfo {author} {\bibfnamefont
  {D.}~\bibnamefont {Kip}}, \ and\ \bibinfo {author} {\bibfnamefont
  {T.}~\bibnamefont {Laarmann}},\ }\href {\doibase 10.1038/ncomms15626}
  {\bibfield  {journal} {\bibinfo  {journal} {Nat. Commun.}\ }\textbf {\bibinfo
  {volume} {8}},\ \bibinfo {pages} {15626} (\bibinfo {year}
  {2017})}\BibitemShut {NoStop}%
\bibitem [{\citenamefont {Zener}(1934)}]{Zener1934}%
  \BibitemOpen
  \bibfield  {author} {\bibinfo {author} {\bibfnamefont {C.}~\bibnamefont
  {Zener}},\ }\href {\doibase 10.1098/rspa.1934.0116} {\bibfield  {journal}
  {\bibinfo  {journal} {Proc. R. Soc. A Math. Phys. Eng. Sci.}\ }\textbf
  {\bibinfo {volume} {145}},\ \bibinfo {pages} {523} (\bibinfo {year}
  {1934})}\BibitemShut {NoStop}%
\bibitem [{\citenamefont {Sillanp{\"{a}}{\"{a}}}\ \emph
  {et~al.}(2006)\citenamefont {Sillanp{\"{a}}{\"{a}}}, \citenamefont
  {Lehtinen}, \citenamefont {Paila}, \citenamefont {Makhlin},\ and\
  \citenamefont {Hakonen}}]{Sillanpaa2006}%
  \BibitemOpen
  \bibfield  {author} {\bibinfo {author} {\bibfnamefont {M.}~\bibnamefont
  {Sillanp{\"{a}}{\"{a}}}}, \bibinfo {author} {\bibfnamefont {T.}~\bibnamefont
  {Lehtinen}}, \bibinfo {author} {\bibfnamefont {A.}~\bibnamefont {Paila}},
  \bibinfo {author} {\bibfnamefont {Y.}~\bibnamefont {Makhlin}}, \ and\
  \bibinfo {author} {\bibfnamefont {P.}~\bibnamefont {Hakonen}},\ }\href
  {\doibase 10.1103/PhysRevLett.96.187002} {\bibfield  {journal} {\bibinfo
  {journal} {Phys. Rev. Lett.}\ }\textbf {\bibinfo {volume} {96}},\ \bibinfo
  {pages} {187002} (\bibinfo {year} {2006})}\BibitemShut {NoStop}%
\bibitem [{\citenamefont {Kling}\ \emph {et~al.}(2010)\citenamefont {Kling},
  \citenamefont {Salger}, \citenamefont {Grossert},\ and\ \citenamefont
  {Weitz}}]{Kling2010}%
  \BibitemOpen
  \bibfield  {author} {\bibinfo {author} {\bibfnamefont {S.}~\bibnamefont
  {Kling}}, \bibinfo {author} {\bibfnamefont {T.}~\bibnamefont {Salger}},
  \bibinfo {author} {\bibfnamefont {C.}~\bibnamefont {Grossert}}, \ and\
  \bibinfo {author} {\bibfnamefont {M.}~\bibnamefont {Weitz}},\ }\href
  {\doibase 10.1103/PhysRevLett.105.215301} {\bibfield  {journal} {\bibinfo
  {journal} {Phys. Rev. Lett.}\ }\textbf {\bibinfo {volume} {105}},\ \bibinfo
  {pages} {215301} (\bibinfo {year} {2010})}\BibitemShut {NoStop}%
\bibitem [{\citenamefont {Zenesini}\ \emph {et~al.}(2009)\citenamefont
  {Zenesini}, \citenamefont {Lignier}, \citenamefont {Tayebirad}, \citenamefont
  {Radogostowicz}, \citenamefont {Ciampini}, \citenamefont {Mannella},
  \citenamefont {Wimberger}, \citenamefont {Morsch},\ and\ \citenamefont
  {Arimondo}}]{zenesini_time-resolved_2009}%
  \BibitemOpen
  \bibfield  {author} {\bibinfo {author} {\bibfnamefont {A.}~\bibnamefont
  {Zenesini}}, \bibinfo {author} {\bibfnamefont {H.}~\bibnamefont {Lignier}},
  \bibinfo {author} {\bibfnamefont {G.}~\bibnamefont {Tayebirad}}, \bibinfo
  {author} {\bibfnamefont {J.}~\bibnamefont {Radogostowicz}}, \bibinfo {author}
  {\bibfnamefont {D.}~\bibnamefont {Ciampini}}, \bibinfo {author}
  {\bibfnamefont {R.}~\bibnamefont {Mannella}}, \bibinfo {author}
  {\bibfnamefont {S.}~\bibnamefont {Wimberger}}, \bibinfo {author}
  {\bibfnamefont {O.}~\bibnamefont {Morsch}}, \ and\ \bibinfo {author}
  {\bibfnamefont {E.}~\bibnamefont {Arimondo}},\ }\href {\doibase
  10.1103/PhysRevLett.103.090403} {\bibfield  {journal} {\bibinfo  {journal}
  {Phys. Rev. Lett.}\ }\textbf {\bibinfo {volume} {103}},\ \bibinfo {pages}
  {090403} (\bibinfo {year} {2009})}\BibitemShut {NoStop}%
\bibitem [{\citenamefont {Trompeter}\ \emph
  {et~al.}(2006{\natexlab{a}})\citenamefont {Trompeter}, \citenamefont
  {Pertsch}, \citenamefont {Lederer}, \citenamefont {Michaelis}, \citenamefont
  {Streppel}, \citenamefont {Br\"auer},\ and\ \citenamefont
  {Peschel}}]{trompeter_visual_2006}%
  \BibitemOpen
  \bibfield  {author} {\bibinfo {author} {\bibfnamefont {H.}~\bibnamefont
  {Trompeter}}, \bibinfo {author} {\bibfnamefont {T.}~\bibnamefont {Pertsch}},
  \bibinfo {author} {\bibfnamefont {F.}~\bibnamefont {Lederer}}, \bibinfo
  {author} {\bibfnamefont {D.}~\bibnamefont {Michaelis}}, \bibinfo {author}
  {\bibfnamefont {U.}~\bibnamefont {Streppel}}, \bibinfo {author}
  {\bibfnamefont {A.}~\bibnamefont {Br\"auer}}, \ and\ \bibinfo {author}
  {\bibfnamefont {U.}~\bibnamefont {Peschel}},\ }\href {\doibase
  10.1103/PhysRevLett.96.023901} {\bibfield  {journal} {\bibinfo  {journal}
  {Phys. Rev. Lett.}\ }\textbf {\bibinfo {volume} {96}},\ \bibinfo {pages}
  {023901} (\bibinfo {year} {2006}{\natexlab{a}})}\BibitemShut {NoStop}%
\bibitem [{\citenamefont {Shevchenko}\ \emph {et~al.}(2010)\citenamefont
  {Shevchenko}, \citenamefont {Ashhab},\ and\ \citenamefont
  {Nori}}]{Shevchenko2010}%
  \BibitemOpen
  \bibfield  {author} {\bibinfo {author} {\bibfnamefont {S.~N.}\ \bibnamefont
  {Shevchenko}}, \bibinfo {author} {\bibfnamefont {S.}~\bibnamefont {Ashhab}},
  \ and\ \bibinfo {author} {\bibfnamefont {F.}~\bibnamefont {Nori}},\ }\href
  {\doibase 10.1016/j.physrep.2010.03.002} {\bibfield  {journal} {\bibinfo
  {journal} {Phys. Rep.}\ }\textbf {\bibinfo {volume} {492}},\ \bibinfo {pages}
  {1} (\bibinfo {year} {2010})}\BibitemShut {NoStop}%
\bibitem [{\citenamefont {Ji}\ \emph {et~al.}(2003)\citenamefont {Ji},
  \citenamefont {Chung}, \citenamefont {Sprinzak}, \citenamefont {Heiblum},\
  and\ \citenamefont {Mahalu}}]{Ji2003}%
  \BibitemOpen
  \bibfield  {author} {\bibinfo {author} {\bibfnamefont {Y.}~\bibnamefont
  {Ji}}, \bibinfo {author} {\bibfnamefont {Y.}~\bibnamefont {Chung}}, \bibinfo
  {author} {\bibfnamefont {D.}~\bibnamefont {Sprinzak}}, \bibinfo {author}
  {\bibfnamefont {M.}~\bibnamefont {Heiblum}}, \ and\ \bibinfo {author}
  {\bibfnamefont {D.}~\bibnamefont {Mahalu}},\ }\href {\doibase
  10.1038/nature01492.1.} {\bibfield  {journal} {\bibinfo  {journal} {Nature}\
  }\textbf {\bibinfo {volume} {422}},\ \bibinfo {pages} {415} (\bibinfo {year}
  {2003})}\BibitemShut {NoStop}%
\bibitem [{\citenamefont {Oliver}(2005)}]{Oliver2005}%
  \BibitemOpen
  \bibfield  {author} {\bibinfo {author} {\bibfnamefont {W.~D.}\ \bibnamefont
  {Oliver}},\ }\href {\doibase 10.1126/science.1119678} {\bibfield  {journal}
  {\bibinfo  {journal} {Science}\ }\textbf {\bibinfo {volume} {310}},\ \bibinfo
  {pages} {1653} (\bibinfo {year} {2005})}\BibitemShut {NoStop}%
\bibitem [{\citenamefont {Trompeter}\ \emph
  {et~al.}(2006{\natexlab{b}})\citenamefont {Trompeter}, \citenamefont
  {Krolikowski}, \citenamefont {Neshev}, \citenamefont {Desyatnikov},
  \citenamefont {Sukhorukov}, \citenamefont {Kivshar}, \citenamefont {Pertsch},
  \citenamefont {Peschel},\ and\ \citenamefont
  {Lederer}}]{trompeter_bloch_2006}%
  \BibitemOpen
  \bibfield  {author} {\bibinfo {author} {\bibfnamefont {H.}~\bibnamefont
  {Trompeter}}, \bibinfo {author} {\bibfnamefont {W.}~\bibnamefont
  {Krolikowski}}, \bibinfo {author} {\bibfnamefont {D.~N.}\ \bibnamefont
  {Neshev}}, \bibinfo {author} {\bibfnamefont {A.~S.}\ \bibnamefont
  {Desyatnikov}}, \bibinfo {author} {\bibfnamefont {A.~A.}\ \bibnamefont
  {Sukhorukov}}, \bibinfo {author} {\bibfnamefont {Y.~S.}\ \bibnamefont
  {Kivshar}}, \bibinfo {author} {\bibfnamefont {T.}~\bibnamefont {Pertsch}},
  \bibinfo {author} {\bibfnamefont {U.}~\bibnamefont {Peschel}}, \ and\
  \bibinfo {author} {\bibfnamefont {F.}~\bibnamefont {Lederer}},\ }\href
  {\doibase 10.1103/PhysRevLett.96.053903} {\bibfield  {journal} {\bibinfo
  {journal} {Phys. Rev. Lett.}\ }\textbf {\bibinfo {volume} {96}},\ \bibinfo
  {pages} {053903} (\bibinfo {year} {2006}{\natexlab{b}})}\BibitemShut
  {NoStop}%
\bibitem [{\citenamefont {Cao}\ \emph {et~al.}(2013)\citenamefont {Cao},
  \citenamefont {Li}, \citenamefont {Tu}, \citenamefont {Wang}, \citenamefont
  {Zhou}, \citenamefont {Xiao}, \citenamefont {Guo}, \citenamefont {Jiang},\
  and\ \citenamefont {Guo}}]{Cao2013}%
  \BibitemOpen
  \bibfield  {author} {\bibinfo {author} {\bibfnamefont {G.}~\bibnamefont
  {Cao}}, \bibinfo {author} {\bibfnamefont {H.-O.}\ \bibnamefont {Li}},
  \bibinfo {author} {\bibfnamefont {T.}~\bibnamefont {Tu}}, \bibinfo {author}
  {\bibfnamefont {L.}~\bibnamefont {Wang}}, \bibinfo {author} {\bibfnamefont
  {C.}~\bibnamefont {Zhou}}, \bibinfo {author} {\bibfnamefont {M.}~\bibnamefont
  {Xiao}}, \bibinfo {author} {\bibfnamefont {G.-C.}\ \bibnamefont {Guo}},
  \bibinfo {author} {\bibfnamefont {H.-W.}\ \bibnamefont {Jiang}}, \ and\
  \bibinfo {author} {\bibfnamefont {G.-P.}\ \bibnamefont {Guo}},\ }\href@noop
  {} {\bibfield  {journal} {\bibinfo  {journal} {Nat. Comms.}\ }\textbf
  {\bibinfo {volume} {{4}}} (\bibinfo {year} {{2013}})}\BibitemShut {NoStop}%
\bibitem [{\citenamefont {Wei}\ \emph {et~al.}(2017)\citenamefont {Wei},
  \citenamefont {van~der Sar}, \citenamefont {Sanchez-Yamagishi}, \citenamefont
  {Watanabe}, \citenamefont {Taniguchi}, \citenamefont {Jarillo-Herrero},
  \citenamefont {Halperin},\ and\ \citenamefont {Yacoby}}]{Wei2017}%
  \BibitemOpen
  \bibfield  {author} {\bibinfo {author} {\bibfnamefont {D.~S.}\ \bibnamefont
  {Wei}}, \bibinfo {author} {\bibfnamefont {T.}~\bibnamefont {van~der Sar}},
  \bibinfo {author} {\bibfnamefont {J.~D.}\ \bibnamefont {Sanchez-Yamagishi}},
  \bibinfo {author} {\bibfnamefont {K.}~\bibnamefont {Watanabe}}, \bibinfo
  {author} {\bibfnamefont {T.}~\bibnamefont {Taniguchi}}, \bibinfo {author}
  {\bibfnamefont {P.}~\bibnamefont {Jarillo-Herrero}}, \bibinfo {author}
  {\bibfnamefont {B.~I.}\ \bibnamefont {Halperin}}, \ and\ \bibinfo {author}
  {\bibfnamefont {A.}~\bibnamefont {Yacoby}},\ }\href {\doibase
  10.1126/sciadv.1700600} {\bibfield  {journal} {\bibinfo  {journal} {Sci.
  Adv.}\ }\textbf {\bibinfo {volume} {3}},\ \bibinfo {pages} {e1700600}
  (\bibinfo {year} {2017})}\BibitemShut {NoStop}%
\bibitem [{\citenamefont {Abanin}\ \emph {et~al.}(2013)\citenamefont {Abanin},
  \citenamefont {Kitagawa}, \citenamefont {Bloch},\ and\ \citenamefont
  {Demler}}]{Abanin2013}%
  \BibitemOpen
  \bibfield  {author} {\bibinfo {author} {\bibfnamefont {D.~A.}\ \bibnamefont
  {Abanin}}, \bibinfo {author} {\bibfnamefont {T.}~\bibnamefont {Kitagawa}},
  \bibinfo {author} {\bibfnamefont {I.}~\bibnamefont {Bloch}}, \ and\ \bibinfo
  {author} {\bibfnamefont {E.}~\bibnamefont {Demler}},\ }\href {\doibase
  10.1103/PhysRevLett.110.165304} {\bibfield  {journal} {\bibinfo  {journal}
  {Phys. Rev. Lett.}\ }\textbf {\bibinfo {volume} {110}},\ \bibinfo {pages}
  {165304} (\bibinfo {year} {2013})}\BibitemShut {NoStop}%
\bibitem [{\citenamefont {Grusdt}\ \emph {et~al.}(2016)\citenamefont {Grusdt},
  \citenamefont {Yao}, \citenamefont {Abanin}, \citenamefont {Fleischhauer},\
  and\ \citenamefont {Demler}}]{Grusdt2016}%
  \BibitemOpen
  \bibfield  {author} {\bibinfo {author} {\bibfnamefont {F.}~\bibnamefont
  {Grusdt}}, \bibinfo {author} {\bibfnamefont {N.~Y.}\ \bibnamefont {Yao}},
  \bibinfo {author} {\bibfnamefont {D.}~\bibnamefont {Abanin}}, \bibinfo
  {author} {\bibfnamefont {M.}~\bibnamefont {Fleischhauer}}, \ and\ \bibinfo
  {author} {\bibfnamefont {E.}~\bibnamefont {Demler}},\ }\href {\doibase
  10.1038/ncomms11994} {\bibfield  {journal} {\bibinfo  {journal} {Nat.
  Commun.}\ }\textbf {\bibinfo {volume} {7}},\ \bibinfo {pages} {11994}
  (\bibinfo {year} {2016})}\BibitemShut {NoStop}%
\bibitem [{\citenamefont {Landau}\ and\ \citenamefont
  {Lifshitz}(2013)}]{landau2013quantum}%
  \BibitemOpen
  \bibfield  {author} {\bibinfo {author} {\bibfnamefont {L.~D.}\ \bibnamefont
  {Landau}}\ and\ \bibinfo {author} {\bibfnamefont {E.~M.}\ \bibnamefont
  {Lifshitz}},\ }\href@noop {} {\emph {\bibinfo {title} {Quantum mechanics:
  non-relativistic theory}}},\ Vol.~\bibinfo {volume} {3}\ (\bibinfo
  {publisher} {Elsevier},\ \bibinfo {year} {2013})\BibitemShut {NoStop}%
\bibitem [{\citenamefont {Wu}\ \emph {et~al.}(2015)\citenamefont {Wu},
  \citenamefont {Ghimire}, \citenamefont {Reis}, \citenamefont {Schafer},\ and\
  \citenamefont {Gaarde}}]{Wu2015}%
  \BibitemOpen
  \bibfield  {author} {\bibinfo {author} {\bibfnamefont {M.}~\bibnamefont
  {Wu}}, \bibinfo {author} {\bibfnamefont {S.}~\bibnamefont {Ghimire}},
  \bibinfo {author} {\bibfnamefont {D.~A.}\ \bibnamefont {Reis}}, \bibinfo
  {author} {\bibfnamefont {K.~J.}\ \bibnamefont {Schafer}}, \ and\ \bibinfo
  {author} {\bibfnamefont {M.~B.}\ \bibnamefont {Gaarde}},\ }\href {\doibase
  10.1103/PhysRevA.91.043839} {\bibfield  {journal} {\bibinfo  {journal} {Phys.
  Rev. A}\ }\textbf {\bibinfo {volume} {91}},\ \bibinfo {pages} {043839}
  (\bibinfo {year} {2015})}\BibitemShut {NoStop}%
\bibitem [{\citenamefont {Jin}\ \emph {et~al.}(2018)\citenamefont {Jin},
  \citenamefont {Xiao}, \citenamefont {Liang}, \citenamefont {Wang},
  \citenamefont {Chen}, \citenamefont {Gong},\ and\ \citenamefont
  {Peng}}]{JinPRA}%
  \BibitemOpen
  \bibfield  {author} {\bibinfo {author} {\bibfnamefont {J.-Z.}\ \bibnamefont
  {Jin}}, \bibinfo {author} {\bibfnamefont {X.-R.}\ \bibnamefont {Xiao}},
  \bibinfo {author} {\bibfnamefont {H.}~\bibnamefont {Liang}}, \bibinfo
  {author} {\bibfnamefont {M.-X.}\ \bibnamefont {Wang}}, \bibinfo {author}
  {\bibfnamefont {S.-G.}\ \bibnamefont {Chen}}, \bibinfo {author}
  {\bibfnamefont {Q.}~\bibnamefont {Gong}}, \ and\ \bibinfo {author}
  {\bibfnamefont {L.-Y.}\ \bibnamefont {Peng}},\ }\href@noop {} {\bibfield
  {journal} {\bibinfo  {journal} {Phys. Rev. A}\ }\textbf {\bibinfo {volume}
  {97}},\ \bibinfo {pages} {043420} (\bibinfo {year} {2018})}\BibitemShut
  {NoStop}%
\bibitem [{\citenamefont {Chini}\ \emph {et~al.}(2014)\citenamefont {Chini},
  \citenamefont {Wang}, \citenamefont {Cheng}, \citenamefont {Wang},
  \citenamefont {Wu}, \citenamefont {Cunningham}, \citenamefont {Li},
  \citenamefont {Heslar}, \citenamefont {Telnov}, \citenamefont {Chu},\ and\
  \citenamefont {Chang}}]{Chang2014}%
  \BibitemOpen
  \bibfield  {author} {\bibinfo {author} {\bibfnamefont {M.}~\bibnamefont
  {Chini}}, \bibinfo {author} {\bibfnamefont {X.}~\bibnamefont {Wang}},
  \bibinfo {author} {\bibfnamefont {Y.}~\bibnamefont {Cheng}}, \bibinfo
  {author} {\bibfnamefont {H.}~\bibnamefont {Wang}}, \bibinfo {author}
  {\bibfnamefont {Y.}~\bibnamefont {Wu}}, \bibinfo {author} {\bibfnamefont
  {E.}~\bibnamefont {Cunningham}}, \bibinfo {author} {\bibfnamefont {P.-C.}\
  \bibnamefont {Li}}, \bibinfo {author} {\bibfnamefont {J.}~\bibnamefont
  {Heslar}}, \bibinfo {author} {\bibfnamefont {D.~A.}\ \bibnamefont {Telnov}},
  \bibinfo {author} {\bibfnamefont {S.-I.}\ \bibnamefont {Chu}}, \ and\
  \bibinfo {author} {\bibfnamefont {Z.}~\bibnamefont {Chang}},\ }\href@noop {}
  {\bibfield  {journal} {\bibinfo  {journal} {Nat. Photonics}\ }\textbf
  {\bibinfo {volume} {8}},\ \bibinfo {pages} {437} (\bibinfo {year}
  {2014})}\BibitemShut {NoStop}%
\bibitem [{\citenamefont {Xiong}\ \emph {et~al.}(2014)\citenamefont {Xiong},
  \citenamefont {Geng}, \citenamefont {Tang}, \citenamefont {Peng},\ and\
  \citenamefont {Gong}}]{Xiong2014}%
  \BibitemOpen
  \bibfield  {author} {\bibinfo {author} {\bibfnamefont {W.-H.}\ \bibnamefont
  {Xiong}}, \bibinfo {author} {\bibfnamefont {J.-W.}\ \bibnamefont {Geng}},
  \bibinfo {author} {\bibfnamefont {J.-Y.}\ \bibnamefont {Tang}}, \bibinfo
  {author} {\bibfnamefont {L.-Y.}\ \bibnamefont {Peng}}, \ and\ \bibinfo
  {author} {\bibfnamefont {Q.}~\bibnamefont {Gong}},\ }\href {\doibase
  10.1103/PhysRevLett.112.233001} {\bibfield  {journal} {\bibinfo  {journal}
  {Phys. Rev. Lett.}\ }\textbf {\bibinfo {volume} {112}},\ \bibinfo {pages}
  {233001} (\bibinfo {year} {2014})}\BibitemShut {NoStop}%
\bibitem [{\citenamefont {Higuchi}\ \emph {et~al.}(2014)\citenamefont
  {Higuchi}, \citenamefont {Stockman},\ and\ \citenamefont
  {Hommelhoff}}]{Higuchi2014}%
  \BibitemOpen
  \bibfield  {author} {\bibinfo {author} {\bibfnamefont {T.}~\bibnamefont
  {Higuchi}}, \bibinfo {author} {\bibfnamefont {M.~I.}\ \bibnamefont
  {Stockman}}, \ and\ \bibinfo {author} {\bibfnamefont {P.}~\bibnamefont
  {Hommelhoff}},\ }\href {\doibase 10.1103/PhysRevLett.113.213901} {\bibfield
  {journal} {\bibinfo  {journal} {Phys. Rev. Lett.}\ }\textbf {\bibinfo
  {volume} {113}},\ \bibinfo {pages} {213901} (\bibinfo {year}
  {2014})}\BibitemShut {NoStop}%
\bibitem [{\citenamefont {Faisal}\ and\ \citenamefont
  {Kami\ifmmode~\acute{n}\else \'{n}\fi{}ski}(1997)}]{FloquetBloch}%
  \BibitemOpen
  \bibfield  {author} {\bibinfo {author} {\bibfnamefont {F.~H.~M.}\
  \bibnamefont {Faisal}}\ and\ \bibinfo {author} {\bibfnamefont {J.~Z.}\
  \bibnamefont {Kami\ifmmode~\acute{n}\else \'{n}\fi{}ski}},\ }\href {\doibase
  10.1103/PhysRevA.56.748} {\bibfield  {journal} {\bibinfo  {journal} {Phys.
  Rev. A}\ }\textbf {\bibinfo {volume} {56}},\ \bibinfo {pages} {748} (\bibinfo
  {year} {1997})}\BibitemShut {NoStop}%
\end{thebibliography}%

\end{document}